\documentclass[12pt,b5paper]{article}
\usepackage{subfigure}
\usepackage{amsmath,epsfig}

\usepackage{graphicx}
\usepackage{dcolumn}
\usepackage{bm}

\begin{document}

\title{$K^{*}$ production in Cu+Cu and Au+Au collisions at
      $\sqrt{s_{\mathrm{NN}}}$= 62.4 GeV and 200 GeV in STAR}

\author {Sadhana Dash(for the STAR Collaboration)\\ 
{\it Institute of Physics ,Bhubaneswar ,INDIA.}\\
e-mail: sadhana@iopb.res.in}
\date{}

\maketitle

\begin{abstract}

We report the measurements of $p_T$ spectra of $K^*$ up to intermediate 
$p_T$ region in mid-rapidity through its hadronic decay channel using the
STAR detector in Au+Au and Cu+Cu collisions at $\sqrt{s_{\mathrm{NN}}}$=
62.4 GeV and 200 GeV. Particle ratios such as $K^{*}/K$ and $K^{*}/\phi$ is
used to understand the rescattering  and regeneration effect on $K^{*}$ 
production in the hadronic medium. The  $K^*$ $v_{2}$ measurement using a
high statistics Au+Au 200 GeV dataset and nuclear modification factor 
measurement supports the quark coalescence model of particle production in
the intermediate $p_T$ range.  

\vspace{0.4in}
\end{abstract}

\section{Introduction}
The main motivation for studying heavy ion collision at high energy
is the study of quantum chromodynamics in extreme conditions of high
temperature and high energy density\cite{whitepaper}.
One of the proposed signatures of a possible phase transition of nuclear 
matter to deconfined state of quarks and gluons is the modification of vector 
meson production rates and their in-medium properties\cite{medium2}.
$K^*$ meson is of particular interest due to its very short life time and its
strange quark content, which makes $K^*$ meson sensitive to the properties
of the dense matter and provide information regarding strangeness production 
from early partonic phase\cite{haibinPRC}\cite{sadhana}. The study of $K^*$ 
meson provides a better understanding on the role of rescattering and 
regeneration effects in hadron production. 
The interplay of these two competing processes is guaged through particle 
ratio studies of $K^{*}/K^{-}$ and $\phi/K^{*}$ in $p+p$ and nucleus-nucleaus
collisions. The mass of $K^*$ is close to the mass of baryons ($p,\Lambda$)
but it is a vector meson, the measurements of $K^*$  nuclear modification 
factor $R_{AA}$ or $R_{CP}$ compared to those of $K_S$ and $\Lambda$ can be 
utilized to distinguish whether the observed differences between the $R_{CP}$ 
of $K_S$ and $\Lambda$ are due to difference in their mass or baryon-meson 
effect\cite{v2scalling}. In the intermediate $p_T$ range,the identified hadron 
elliptic flow $v_{2}$ measurements have shown that the hadronic $v_{2}$ follows
a simple scaling of the number of constituent quarks in the hadrons. The $K^*$
$v_2$ measurement may reveal the $K^*$ production mechanism in hadronic phase\cite{haibinPRC}.

\section{Experiment and Data Analysis}

The results discussed here are taken from Au+Au and Cu+Cu collisions at
$\sqrt{s_{\mathrm{NN}}}$= 200 GeV and $\sqrt{s_{\mathrm{NN}}}$= 62.4 GeV at
RHIC. The Time Projection Chamber\cite{tpc} within STAR was used to measure
the $K^*$ production via its hadronic decay channel.
The unlike sign K$\pi$ invariant mass distribution was reconstructed eventwise
from random combination of K$\pi$ pairs. The combinatorial background 
distribution was built by using mixed-event technique\cite{eventmix}.
The mixed event generated was normalized to subtract the background in the
same event unlike-sign invariant mass spectrum. The $K^{*}$ signal was observed
after subtracting the normalized mixed event background from the unlike sign 
spectrum.

\begin{figure*}[htp]
\begin{minipage}{0.4\textwidth}
\centering
\includegraphics[height=13pc,width=14pc]{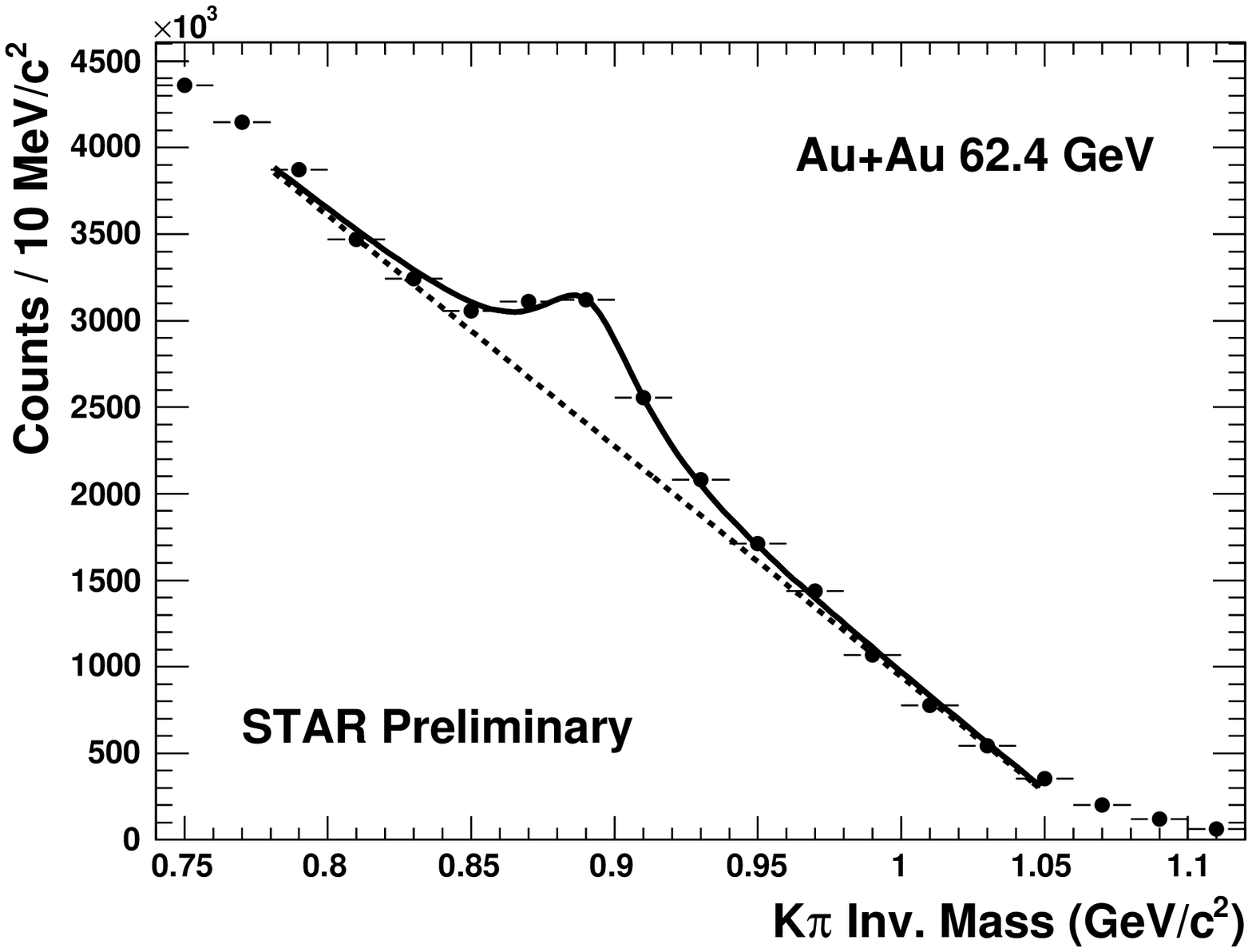}
\end{minipage}
\begin{minipage}{0.4\textwidth}
\centering
\includegraphics[height=13pc,width=14pc]{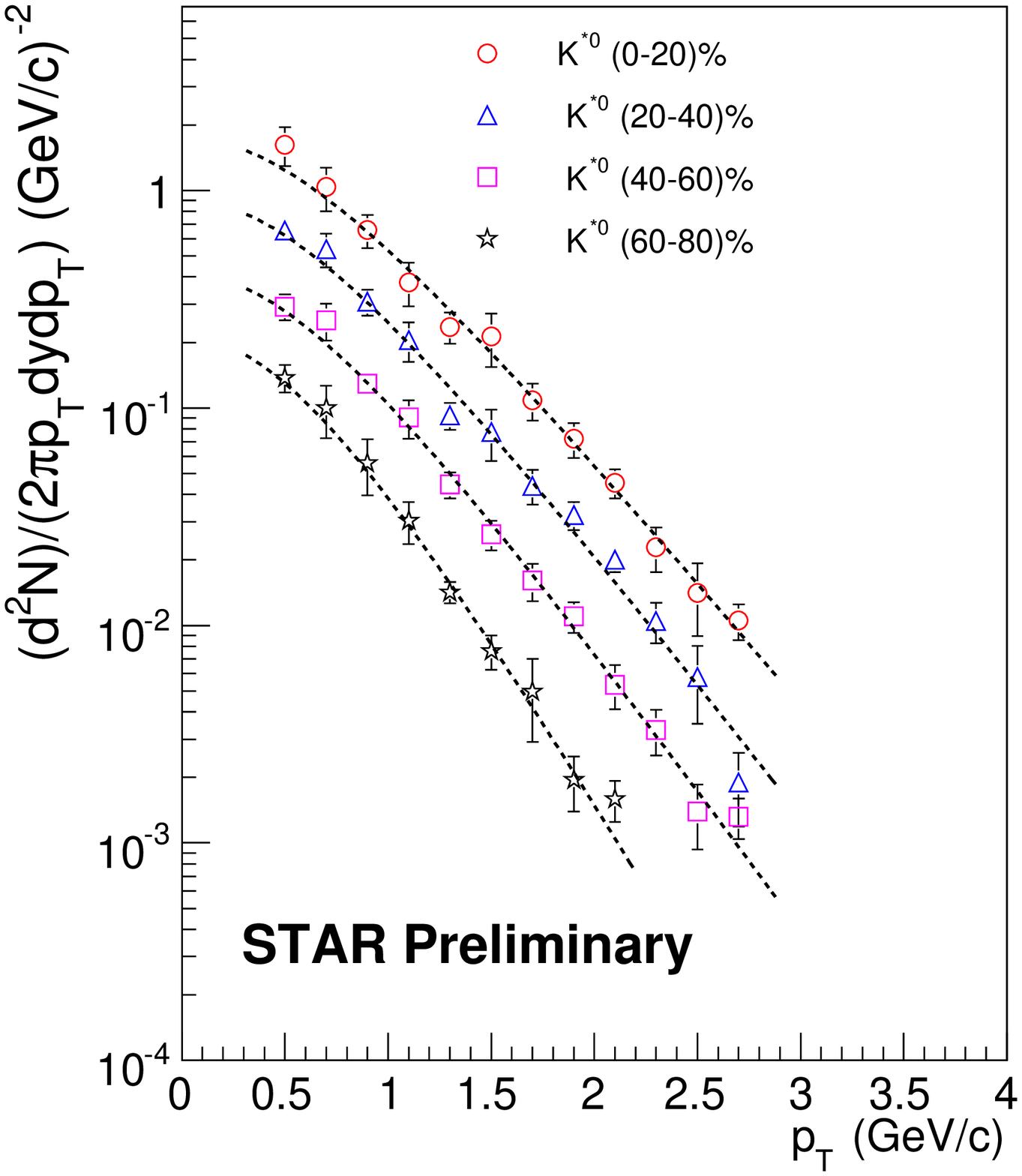}
\end{minipage}
\caption{The K$\pi$ pair invariant mass spectrum after mixed-event
background subtraction fitted to SBW + RB(Left panel) and $p_{T}$ spectra in 
Au+Au collisions at 62.4 GeV fitted to an exponential function(Right panel).}
\label{signal}
\end{figure*}

\section{Results}

\begin{figure*}[htp]
\begin{minipage}{0.45\textwidth}
\centering
\includegraphics[height=13pc,width=14pc]{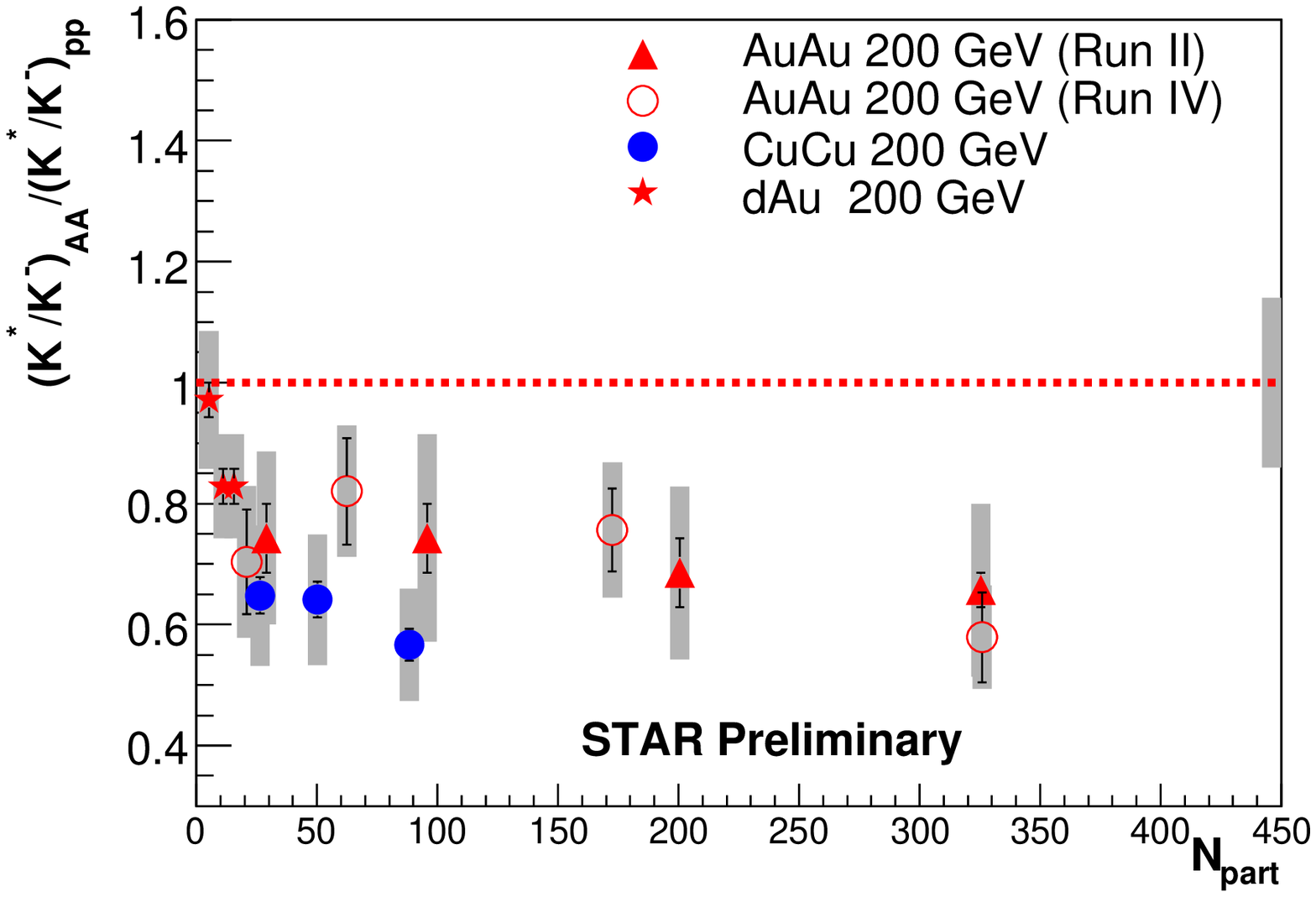}
\end{minipage}
\begin{minipage}{0.45\textwidth}
\centering
\includegraphics[height=13pc,width=14pc]{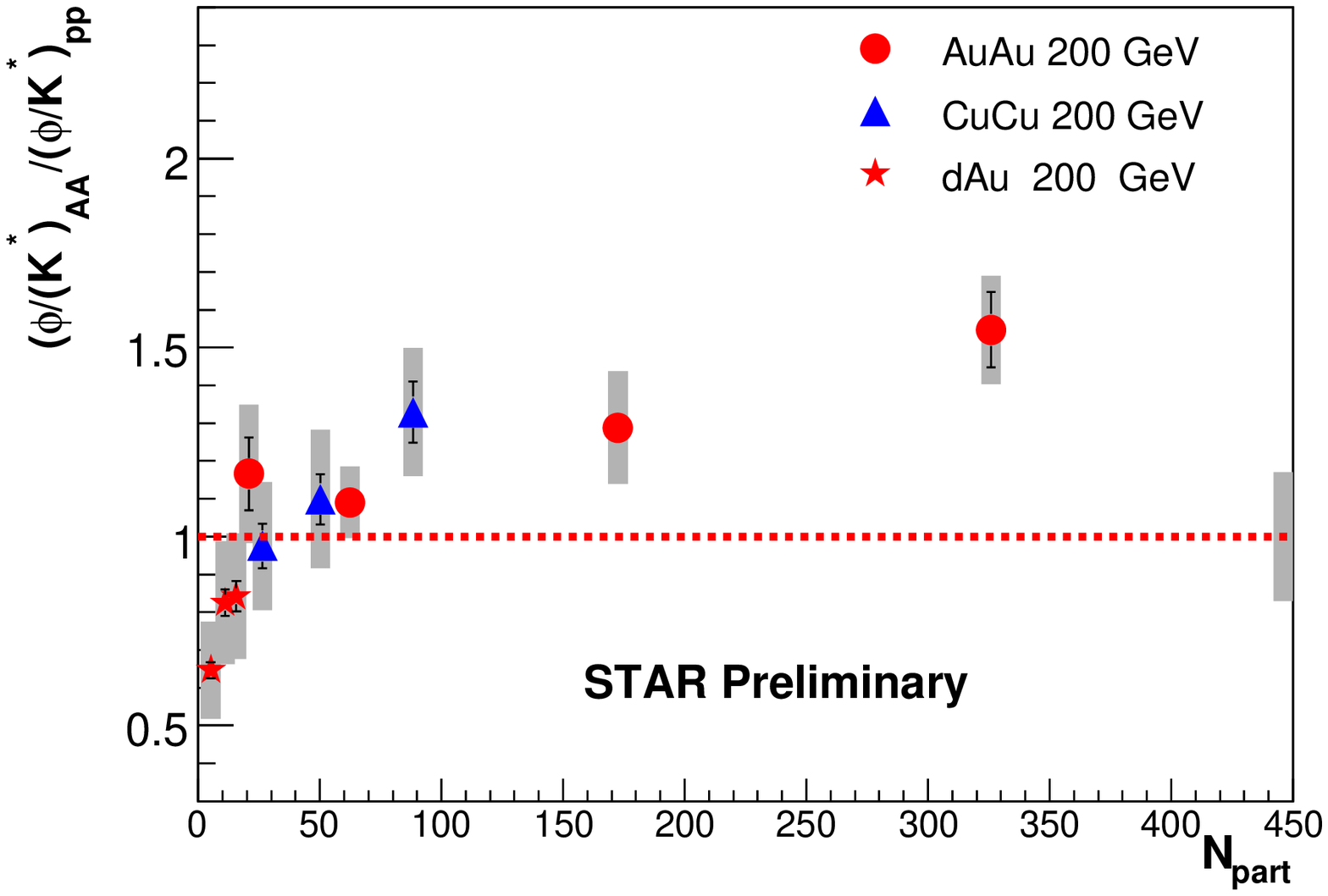}
\end{minipage}
\caption{$K^{*0}/K^{-}$ ratio (left panel)and $\phi/K^{*0}$ ratio(right panel)
normalized by their values in p+p collisions at 200 GeV as a function of number
of participants.The error bars correspond to statistical errors whereas the 
error bands are systematic uncertainities.The width of band(right)at value 1
on y-axis represents the error in $K^{*0}/K^{-}$ in $p+p$ collisions.}
\label{particleratio}
\end{figure*}

Figure \ref{signal}(left panel) shows the unlike sign K$\pi$ invariant mass
spectrum after normalized mixed event background subtraction in minimum bias
Au+Au collisions at $\sqrt{s_{\mathrm{NN}}}$= 62.4 GeV. The invariant mass 
distribution is fitted to the function SBW + RBG where SBW is the non 
relativistic Breit-Wigner function and RBG is the linear function describing 
the residual background\cite{haibinPRC}.
Figure \ref{signal}(right panel) shows the acceptance and efficiency corrected
$K^{*}$ $p_{T}$ spectra for $|y|<0.5$ for different centralities in Au+Au 
collisions at $\sqrt{s_{\mathrm{NN}}}$= 62.4 GeV.
Figure \ref{particleratio}(left panel) depicts the $K^{*0}/K^{-}$ ratio
normalized by their values in p+p collisions at the same beam energy.
The decrease of the ratio with the number of participants indicates that
the rescattering of the decay particles dominates over resonance regeneration.
In Figure \ref{particleratio}(right panel), we observe an increase of
$\phi/K^{*0}$ ratio normalized by their values in p+p collisions with number
of participants.This again favours dominance of rescattering effect on $K^{*}$.

\begin{figure*}[htp]
\begin{minipage}{0.45\textwidth}
\centering
\includegraphics[height=13pc,width=14pc]{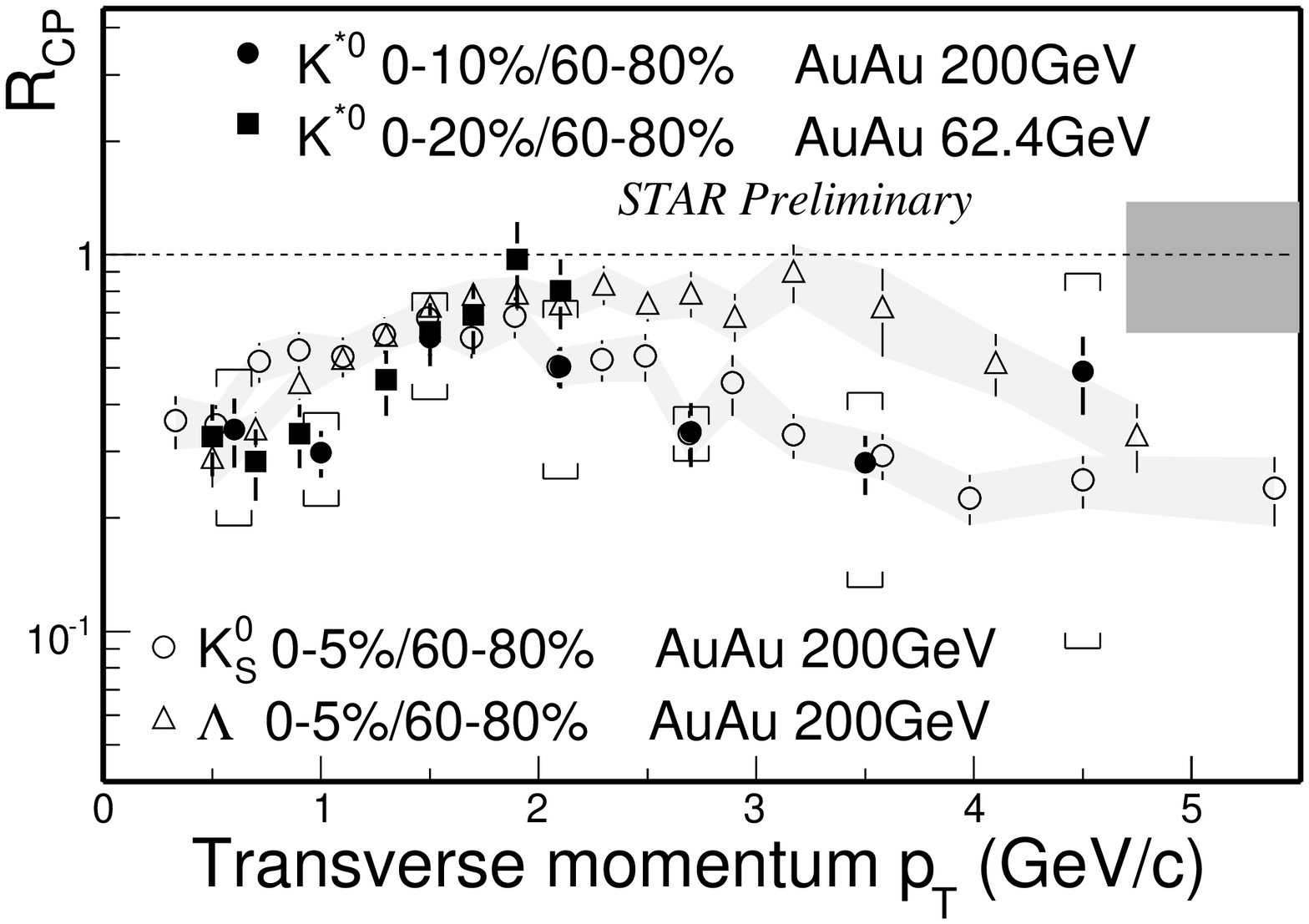}
\end{minipage}
\begin{minipage}{0.45\textwidth}
\centering
\includegraphics[height=13pc,width=14pc]{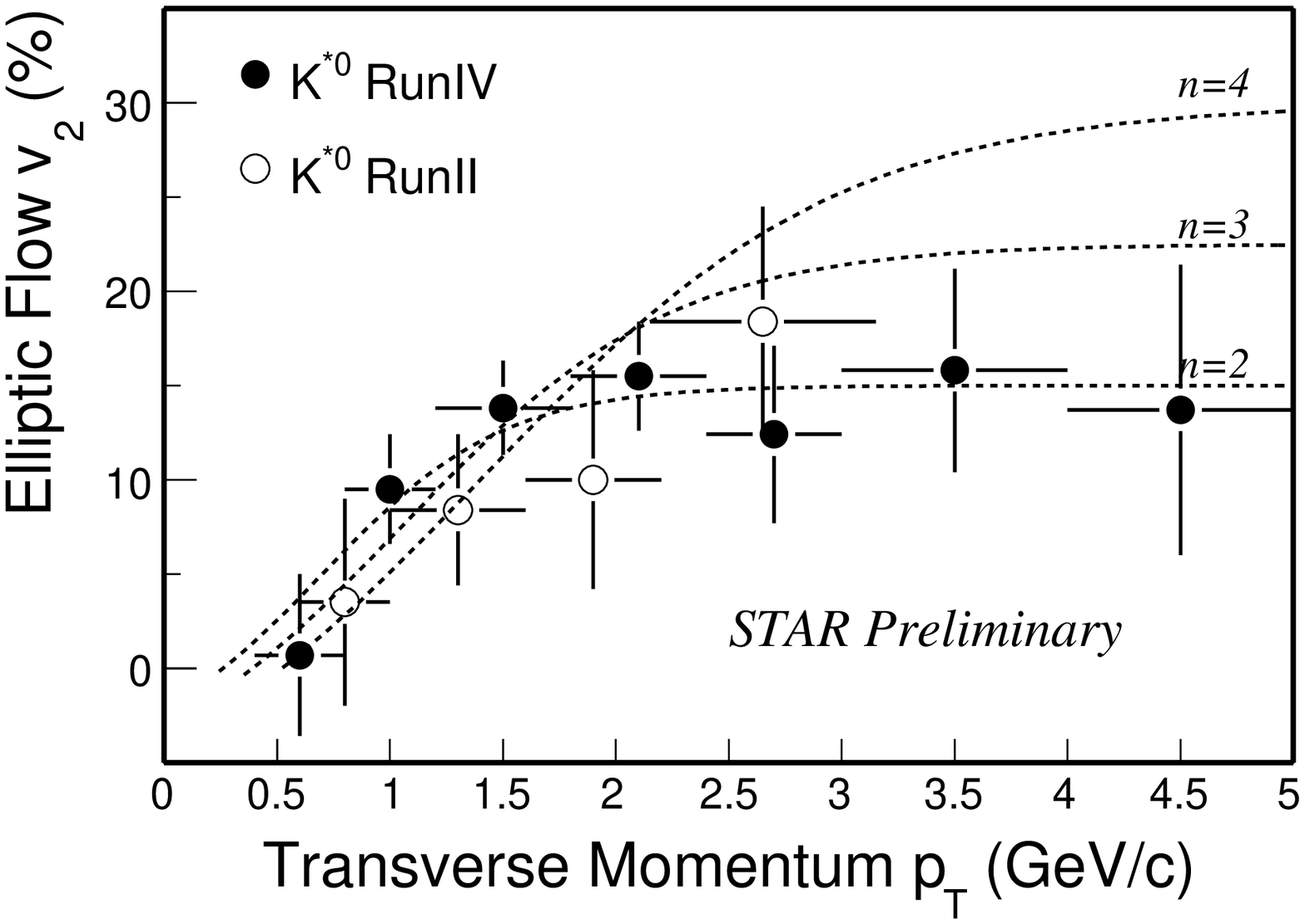}
\end{minipage}
\caption{$K^{*0}$ $R_{CP}$(left panel)as a function of $p_{T}$ in Au+Au 
collisions at 200 GeV and 62.4 GeV compared to $R_{CP}$ of $K_S^0$, $\Lambda$.
$K^{*0}$ $v_{2}$ as a function of $p_{T}$ in minbias Au+Au collisions at
200 GeV(right panel).}
\label{v2}
\end{figure*}

Figure \ref{v2}(left panel) shows the $K^{*0}$ $R_{CP}$ as function of $p_T$ 
compared to the $\Lambda$ and $K_S^0$ $R_{CP}$. For $p_T <$ 1.8 GeV/c, the
$K^{*0}$ $R_{CP}$ in Au+Au collisions at 200 GeV and 62.4 GeV are smaller
than that of $\Lambda$ and $K_S^0$, indicating strong rescattering of $K^{*0}$
decay particles at low $p_T$. For $p_T >$ 1.8 GeV/c, the $K^{*0}$ $R_{CP}$ in 
Au+Au collisions at 200 GeV is more closer to the $K_S^0$ $R_{CP}$ which
favours a baryon-meson effect of the particle production in the intermediate
$p_T$ region.

Figure \ref{v2} shows $K^{*0}$ elliptic flow $v_2$ as a function of $p_T$ in
minimum bias Au+Au collisions at $\sqrt{s_{\mathrm{NN}}}$= 200 GeV. It was fit
with function\cite{Xinv2}:
\begin{equation}\label{v2func}
v_2(p_T,n) = \frac{an}{1+exp(-(p_T/n-b)/c)}-dn
\end{equation}
where $a,b,c$ and $d$ are fixed parameters extracted by fitting $K_S^0$ and
$\Lambda \ v_2$ data points in reference\cite{Xinv2},and $n$ is an open
parameter representing the number of constituent quarks. Fitting the $K^{*0}$
$v_{2}$ data with function given in Eqn.\ref{v2func} gives a value of 
$n=2.0\pm0.3$ with $\chi^2/ndf$ = 2/6. This indicates that $K^{*}$ are 
dominantly produced from direct quark combinations, and $K^*$ regeneration 
is negligible in the hadronic stage.

\section{Summary}
The preliminary results on the $K^{*}$ production in Au+Au and Cu+Cu collisions
measured with the STAR detector at RHIC at $\sqrt{s_{\mathrm{NN}}}$ = 62.4 GeV
and  $\sqrt{s_{\mathrm{NN}}}$ = 200 GeV are presented.
The particle ratio and $R_{CP}$ measurement supports the dominance of 
rescattering effect over the regeneration mechanism in $K^{*}$ production. 
A significant non-zero  elliptic flow $v_{2}$ of K* is measured using the high
statistics minimum bias Au+Au 200 GeV data. In the intermediate $p_{T}$, the
number of quarks from $v_{2}$ scaling was found to be 2.0 +- 0.3 which implies
that the observed K* is predominantly produced by direct quark combinations.

\end{document}